# NONLINEAR DYNAMICS OF INFECTIOUS DISEASES TRANSFER WITH POSSIBLE APPLICATIONS FOR TUBERCULAR INFECTION

V.D. Krevchik[1,5], T.V. Novikova[2], Yu. I. Dahnovsky[3],

M.B. Semenov[1,5], E.V. Shcherbakova[1], Kenji Yamamoto[4]

[1] Physics Department, Penza State University, Penza 440017, Russia
physics@diamond.stup.ac.ru
[2] Penza regional tubercular prophylactic center, Russia
[3] Department of Physics & Astronomy/3905, 1000 E. University Ave.,
University of Wyoming, Laramie, WY, 82071, USA
[4] Research Institute of International Medical Center, Japan
backen@ri.imcj.go.jp
[5] Institute for Basic Research, P.O. Box 1577, Palm Harbor, FL 34682, USA
ibr@verizon.net

In this paper, we model a nonlinear dynamics of infectious diseases transfer. Particularly, we study possible applications to tubercular infection in models with different profiles (peak values) of the population density dependence on spatial coordinates. Our approach is based on the well known method of instantons which has been used by the authors to describe kinetics of adiabatic chemical reactions as a function of the heat-bath temperature and other system parameters. In our approach, we use "social temperature" $T^*$ as one of the controlling parameters. Increase of $T^*$ leads to acceleration of the infectious diseases transfer. The "blockage" effect for the infectious diseases transfer has been demonstrated in the case when peak values (in the population density) are equal to one and under condition that the "social temperature" is low. Existence of such effect essentially depends from environment "activity" (social and prophylactic). Results of our modeling qualitatively meet the tuberculosis dynamic spread data in Penza region of Russia.

# 1. Introduction

The problem of exact modeling of different types of infectious diseases transfer is of great importance [1, 2, 3, 4]. In most cases existing models are of "macroscopic" character (the scale is countrywide). A number of "macroscopic" models can be described as "network" models, percolation-, reactionary-diffusion-, oscillating, stochastic-, multivariable Markovian models [1], etc. Another type of models are of a microscopic character which are used at the molecular level: mutation models; models, which reveal infectious stability to different medicine influence, etc. Up to date, models of an intermediate (mesoscopic) level practically are not developed. Such models could give important (integrating) information about infectious "carrier" (like incubation period or latency). Also, such models would allow investigation of features of controllable infectious transfer at the scale of connecting settlements, e.g., towns.

Existing spectrum of the infectious transfer models has been discussed at the "International Symposium on Transmission Models for Infectious Diseases" [2, 3, 4] at the Tokyo International Medical Center of Japan, and vital necessity of the mesoscopic models for infectious transfer has been convincingly demonstrated. Although the most reports at this Conference has been devoted to modeling of the supposed global pandemic influenza (with account for high risks and for the infectious transfer high rates); the mesoscopic model for the tuberculosis transfer has been recognized as important one as well. Importance of the research at this level can be also confirmed by the integrated statistics for the tuberculosis diseases growth in Russia since 1991 (see Fig. 1). This tendency is kept today in Russia, and problem is complicated because of high population decrease (see Fig. 2).

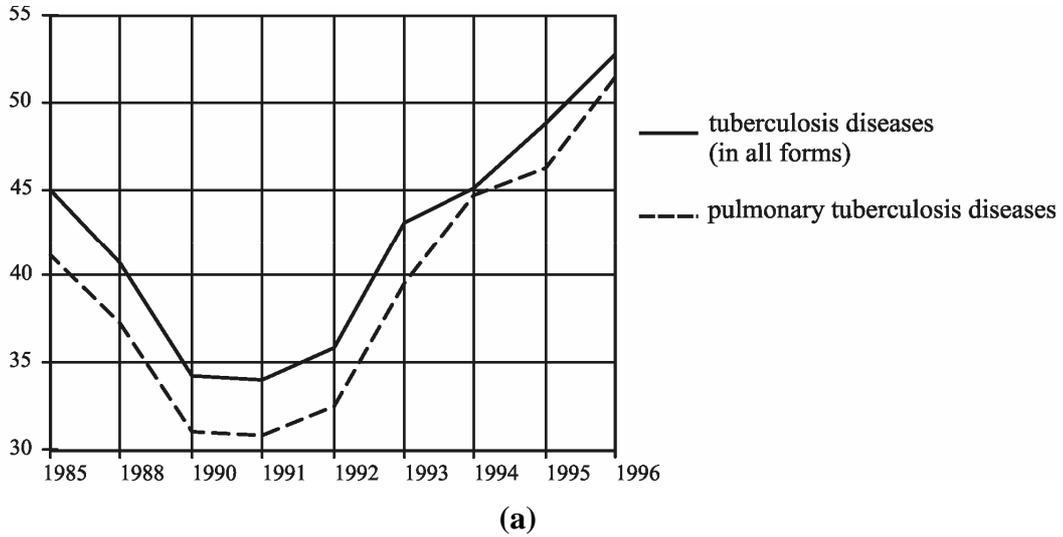

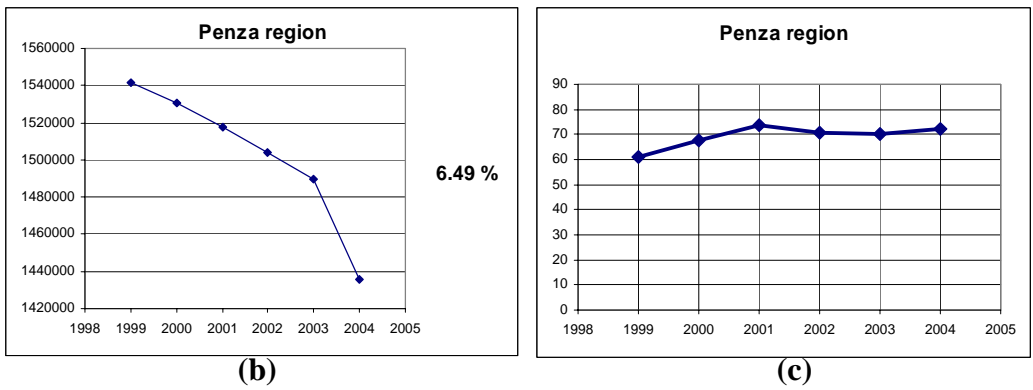

**Fig. 1.** (a) Tendencies for the tuberculosis diseases growth in Russia since 1991; (b) Population decrease in Penza region, Russia; (c) Tuberculosis diseases growth in Penza region, Russia.

## 2. The model description

Prior to mathematical formulation of the model, we will discuss on some assumptions. For example, with an account for the tuberculosis diseases features (as social diseases), preferential (or probable) directions of infectious transfer are directions from settlements (towns) with lower population density (Fig. 3) to settlements with higher population density. The infection transfer rate essentially depends from social-economic conditions, which we indirectly describe by the "social temperature" parameter. With higher "social temperature", more complicated social-economic conditions are realized (unemployment level increases, inflation increases, medical service for the essential population part is getting worse, etc.); as the result, infectious transfer rate becomes higher. Environment (where infection transfers) essentially influence on the infectious disease transfer dynamics. Particularly, active prophylactics can increase resistibility (or "dissipation") of environment to the infectious disease transfer. On the other hand, destabilizing or excitation (oscillating) influence on the environment can lead to the infectious transfer rate growth. Besides, "inertia", which is related to incubation period of disease, can be included in the model through the parameter similar to "mass". All the abovementioned factors are considered in suggested mesoscopic model.

To construct the model, we develop and apply well approved instanton method, which has been used earlier in dissipative tunnel chemical kinetics research (with account of the heat bath influence) [5, 6]. For this purpose, we will specify the formulation of transfer (kinetic) problem and will show how Hamiltonian of the initial reactionary system can be transformed to the traditional one, which can describe transfer with account of "dissipation". To make it, we introduce the reaction coordinate (infectious transfer direction); and the potential profile along this coordinate will be of the form shown in Fig. 2.

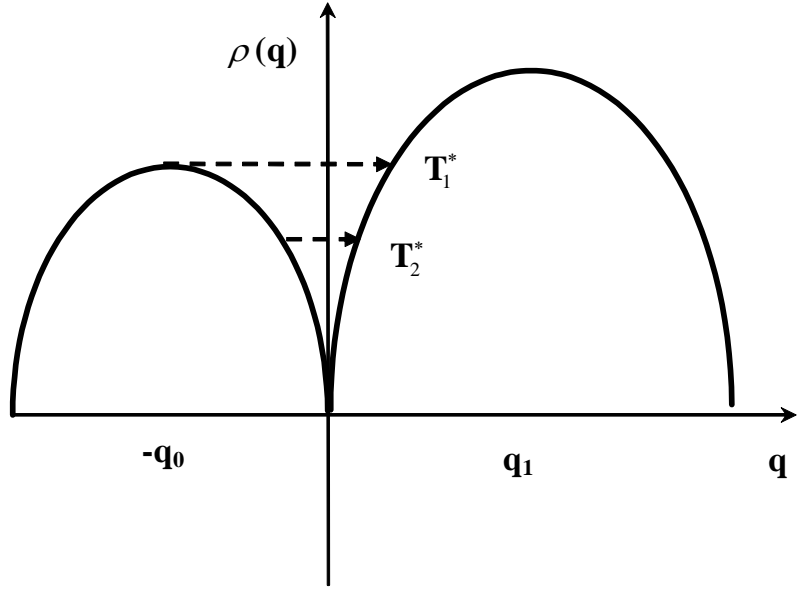

**Fig. 2.** Distribution of an infection (of infectious diseases transfer) along a line connecting two neighbor settlements with various population density $\rho(q)$, depending on "social temperature" $T^*$ ($T_2^* > T_1^*$).

The condition of the reacting system in an environment is determined by the multidimensional potential energy surface. Appropriate approximation to study "low temperature" kinetics at such a surface is the following potential energies:

$$U_i = \sum_{i=1}^{N} \frac{1}{2} \omega_{0i}^{2} (x_i + x_{0i})^2, \quad U_f = \sum_{i=1}^{N} \frac{1}{2} \omega_{0i}^{2} (x_i - x_{0i})^2 - \Delta I. \qquad (1)$$

At low temperatures the potential energy terms of the initial and final states are assumed to be presented as the set of oscillators with frequencies $\omega_{0i}$, relatively shifted to $2x_{oi}$, the oscillator "masses" are chosen to be equal to 1 , hence, their coordinates are renormalized to the square root of the cor-responding mass. The term of the final state lies lower than the initial one by the value of $\Delta I$ (the "reaction heat" or asymmetry parameter in the population density profile).

The plane, being the intersection surface of the two paraboloids, is determined by the equation:

$$2\lambda \sum_{i=1}^{N} \gamma_i x_i = -\Delta I, \tag{2}$$

where

$$\gamma_i = \frac{1}{\lambda} x_{0i} \omega_{0i}^2, \tag{3}$$

$$\lambda^2 = \sum_{i=1}^{N} \omega_{0i}^4 x_{0i}^2 \tag{4}$$

so that the normalization condition is fulfilled,

$$\sum_{i=1}^{N} \gamma_i^2 = 1. \tag{5}$$

The coordinate normal to the plane (2) stands for a transfer coordinate. Let us select it out of the whole coordinate set so that the remaining in the plane (2) $(N-1)$ oscillators are mutually independent but and only linearly connected with the chosen transfer coordinate. For this aim we consider the orthogonal rotation of the coordinate system, so that one coordinate coincides with the transfer one:

$$y_1 = \sum_{i=1}^{N} \gamma_i x_i, \tag{6}$$

whereas the other $(N-1)$ coordinates diagonalize the potential energy in the plane (2):

$$y_j = \sum_{i=1}^{N} U_{ji} x_i, \qquad (7)$$

and $U_{1i} = \gamma_i$. The quadratic form $\sum_{i=1}^{N} \omega_{0i}^{2} x_i^{2}$ may be transformed to

$$\omega_1^2 y_1^2 + 2 y_1 \sum_{\alpha=2}^{N} C_\alpha y_\alpha + \sum_{\alpha=2}^{N} \omega_\alpha^2 y_\alpha^2, \qquad (8)$$

where $\omega_\alpha^2$ satisfies the equation for eigenvalues. To obtain this equation we should diagonalize the quadratic form

$$\sum_{i=1}^{N} \omega_{0i}^{2} x_i^{2}, \qquad (9)$$

Under condition that the transfer coordinate is determined by Eq. (6), and other oscillators coordinates are chosen so that there are not terms of $y_i y_\alpha$ - type $(i, \alpha \geq 2)$, but there are terms of the transfer coordinate $y_1$ interaction with co-ordinates $y_i$ $(i \geq 2)$. For this purpose we should fulfill diagonalization of the quadratic form:

$$\sum_{i=1}^{N} \omega_{0i}^{2} U_{\alpha i} U_{\alpha' i} = \omega_i^2 \delta_{\alpha \alpha'}, \qquad (10)$$

where $U_{\alpha i}$ is element of an orthogonal matrix. Let's multiply both parts of this equation on $U_{\alpha' i}$ and make summation over $\alpha'$ from $2$ to $N$ with an account for the orthogonality condition for the matrix of transformation,

$$U_{\alpha k} = \frac{\gamma_k C_\alpha}{\omega_{0k}^2 - \omega_\alpha^2}, \tag{11}$$

where

$$C_\alpha = \sum_{i=1}^{N} \omega_{0i}^2 \gamma_i U_{\alpha i}. \tag{12}$$

Substituting $U_{\alpha k}$ from (11) to (12), we obtain the equation, which determines the eigenvalues $\omega_\alpha^2$,:

$$\sum_{i=1}^{N} \frac{\omega_{0i}^2 \gamma_i^2}{\omega_{0i}^2 - \omega_\alpha^2} = 1. \tag{13}$$

One eigenvalue $\omega_1^2 = 0$ should be excluded from the consideration. With an account for Eq. (5), the equation (13) can be transformed to following form:

$$\sum_{i=1}^{N} \frac{\gamma_i^2}{\omega_{0i}^2 - \omega_\alpha^2} = 0. \tag{14}$$

Let's define now factors $C_\alpha$ from Eq. (12) and orthogonality conditions for the matrix of transformation:

$$C_\alpha = \left[ \sum_{i=1}^{N} \frac{\gamma_i^2}{\left(\omega_{0i}^2 - \omega_\alpha^2\right)^2} \right]^{-1/2}. \tag{15}$$

Let's note, that

$$\omega_1^2 = \sum_{i=1}^{N} \omega_{0i}^2 \gamma_i^2, \tag{16}$$

Hence, the system Hamiltonian may be written as:

$$\widehat{H} = \frac{p_1^2}{2} + v_1(y_1) + y_1 \sum_{\alpha=2}^{N} C_\alpha y_\alpha + \frac{1}{2} \sum_{\alpha=2}^{N} \left( p_\alpha^2 + \omega_\alpha^2 y_\alpha^2 \right), \tag{17}$$

where

$$v_1(y_1) = \left(\frac{1}{2}\omega_1^2 y_1^2 + \lambda y_1\right)\theta\left(-\frac{\Delta I}{2\lambda} - y_1\right) +$$
$$\left(\frac{1}{2}\omega_1^2 y_1^2 - \lambda y_1 - \Delta I\right)\theta\left(\frac{\Delta I}{2\lambda} + y_1\right).$$
(18)

Probability for the "particle" (infection carrier) transfer can be found in the quasi-classical approximation. We should also assume quasi-stationary transfer. For the case of nonzero temperature we will determine the transfer probability per unit time for an infection carrier as:

$$\Gamma = 2T\frac{\operatorname{Im} Z}{\operatorname{Re} Z}.$$
(19)

Here, $Z$ is a partition function, being the complex value because of the decay. Coherent oscillation [5, 6]), (which we do not consider in the present work) are probable in the case of weak interaction with oscillator environment. To calculate $\Gamma$, it is convenient to represent $Z$ in form of the path integral [5, 6]:

$$Z = \prod_\alpha \int Dy_1 \int Dy_\alpha \exp[-S\{y_1; y_\alpha\}].$$
(20)

Since we are not interested in the initial and final oscillator states, it is possible to integrate over the path $y_\alpha(\tau)$, taking the initial conditions $y_\alpha(-\beta/2) = y_\alpha(\beta/2)$, where $\beta \equiv T^{-1}$ and $T$ is "social temperature". Then the action depends only on the path $y_1(\tau)$:

$$S\{y_1\} = \int_{-\beta/2}^{\beta/2} d\tau \left[ \frac{1}{2} \dot{y}_1^2 + v(y_1) + \frac{1}{2} \int_{-\beta/2}^{\beta/2} d\tau' K(\tau - \tau') y_1(\tau) y_1(\tau') \right],$$

(21)

where

$$v(y_1) = v_1(y_1) - \frac{1}{2} \sum_{\alpha=2}^{N} \frac{C_\alpha^2}{\omega_\alpha^2} y_1^2.$$

(22)

Here, the potential is renormalized, or adiabatic potential is introduced. The kernel of an integral term in (21) depends only on the oscillator environment parameters. Under expanding of a kernel $K(\tau)$ in Fourier series, Fourier-coefficients $\zeta_n$ are defined as:

$$\zeta_n = v_n^2 \sum_{\alpha=2}^{N} \frac{C_\alpha^2}{\omega_\alpha^2 (\omega_\alpha^2 + v_n^2)}.$$

(23)

where $v_n \equiv 2\pi n T$ is the Matsubara's frequency.

Let us perform the following shift of the coordinate $y_1$

$$q = y_1 + \frac{\Delta I}{2\lambda}.$$

(24)

Then

$$v(q) = \frac{1}{2} \omega_0^2 (q + q_0)^2 \theta(-q) + \left[ \frac{1}{2} \omega_0^2 (q - q_1)^2 - \Delta I \right] \theta(q),$$

(25)

where

$$\omega_0^2 = \omega_1^2 - \sum_{\alpha=2}^{N} \frac{C_\alpha^2}{\omega_\alpha^2},$$

$$q_0 = \frac{\lambda}{\omega_0^2} - \frac{\Delta I}{2\lambda}, \qquad q_1 = \frac{\lambda}{\omega_0^2} + \frac{\Delta I}{2\lambda}. \qquad (26)$$

The form of potential (25) with account of "Euclidean turn" in the instanton approach [5, 6] is represented in Fig. 2.

The partition function $Z$ can be calculated in the quasiclassical approximation. There is a trajectory $q_B(\tau)$ which makes the main contribution to the action $S\{q\}$. This trajectory minimizes the action functional and satisfies the Euler-Lagrange equation:

$$-\ddot{q}_B(\tau) + \frac{\partial v(q_B)}{\partial q_B} + \int_{-\beta/2}^{\beta/2} d\tau' K(\tau - \tau') q_B(\tau') = 0, \qquad (27)$$

$q_B(\tau)$ is considered to be a periodic function:

$$q_B(\tau) = q_B(\tau + \beta). \qquad (28)$$

The form $q_B(\tau)$ is defined by the "particle" (infectious carrier) motion in the inverted potential, $-v(q)$. A particle begins its motion (in the zero temperature case) on the top of the potential, $(-v(q))$, i.e. in the point $-q_0$, then it passes through the minimum point, $(q_B = 0)$, at $\tau = -\tau_0$ and attains $q_B = q_0$ (in the symmetric potential case) at $\tau = 0$. Then the particle repeats its trajectory in the reverse order. Such a trajectory stands for an instanton. Notice that the instanton action value does not depend on the position of the instanton centre. The time $\tau_0$ is determined by the equation:

$$q_B(\tau_0) = 0 \qquad (29)$$

The trajectory $q_B(\tau)$ is represented in Fig. 3. The solution of equation (27) is substantially simplified by the introduction of time $\tau_0$, because the coordinate dependent step-functions may be replaced by the respective time dependent ones:

$$\theta(-q_B) = \theta(-\tau_0 - \tau) + \theta(\tau - \tau_0);$$
$$\theta(q_B) = \theta(\tau + \tau_0) - \theta(\tau - \tau_0).$$

(30)

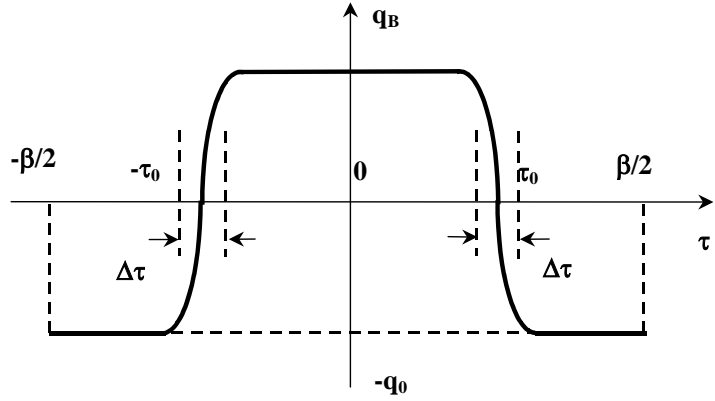

**Fig. 3.** The instanton trajectory $q_B(\tau)$. $\tau_0$ is the instanton centre, $\Delta\tau$ is the instanton width.

Let us expand $q_B(\tau)$ in Fourier series:

$$q_B(\tau) = \beta^{-1} \sum_{n=-\infty}^{\infty} q_n \exp(i\nu_n \tau). \qquad (31)$$

Expanding $\theta$-functions and kernel $K(\tau)$ in Fourier series we obtain the equation determining the Fourier-coefficients $q_n$. This equation may be solved exactly, and we find:

$$q_B(\tau) = -q_0 + \frac{2(q_0+q_1)\tau_0}{\beta} + \frac{2\omega_0^2(q_1+q_0)}{\beta}\sum_{n=1}^{\infty}\frac{\sin\nu_n\tau_0 \cdot \cos\nu_n\tau}{\nu_n\left(\nu_n^2+\omega_0^2+\zeta_n\right)}, \quad (32)$$

where $\nu_n = 2\pi n/\beta$ is Matsubara frequency, and $\zeta_n$ is defined by equation (23). Substituting (32) into the expression for action, we obtain:

$$S_B = 2\omega_0^2(q_0+q_1)q_0\tau_0 - \frac{2\omega_0^2(q_0+q_1)^2\tau_0^2}{\beta} -$$

$$-\frac{4\omega_0^4(q_0+q_1)^2}{\beta}\sum_{n=1}^{\infty}\frac{\sin^2\nu_n\tau_0}{\nu_n^2\left(\nu_n^2+\omega_0^2+\zeta_n\right)}. \quad (33)$$

Thus, quasiclassical action in the one-instanton approximation is analytically found. In the simplest case without interaction with environment, i.e. $\zeta_n=0$, $\tau_0$ is determined from equations (29) and (32):

$$\tau_0 = \frac{1}{2\omega_0}Arcsh\left[\frac{q_0-q_1}{q_0+q_1}sh\frac{\omega_0\beta}{2}\right] + \frac{\beta}{4}. \quad (34)$$

The action may be found from expressions (33) and (34):

$$S_B = \frac{\omega_0(q_1^2-q_0^2)}{2}Arcsh\left[\frac{q_1-q_0}{q_1+q_0}sh\frac{\omega_0\beta}{2}\right] - \frac{\omega_0^2(q_1^2-q_0^2)}{4}\beta +$$

$$+\frac{\omega_0(q_1+q_0)^2}{2}\left\{\frac{\left[ch\frac{\omega_0\beta}{2} - \left[1+\left(\frac{q_1-q_0}{q_1+q_0}\right)^2 sh^2\frac{\omega_0\beta}{2}\right]\right]^{1/2}}{sh\frac{\omega_0\beta}{2}}\right\}.$$

$$(35)$$

In the symmetric case

$$S_B = 2\omega_0 q_0^2 \, th\frac{\omega_0 \beta}{4}. \tag{36}$$

Thus, we have completed construction of our model, and the transfer probability $\Gamma$ (for infectious carrier transfer) with the exponential accuracy can be estimated as $\Gamma \sim \exp(-S)$. Below, we turn to analysis of the results.

### 3. Results and discussion

Dependence of the transfer probability $\Gamma$ (for infectious carrier) on "social temperature" $T^*$ is depicted on Fig. 4. This probability is very sensitive to the local oscillator mode frequency and to the interaction constant with the heat bath (or contact environment), Fig. 5. These results are quite expectative: effectiveness of interaction increases with the oscillator frequency rise that gives corresponding growth in the transfer energy (for infectious carrier) and takes the transfer probability increase ("transition" from curve 1 to curve 2, Fig. 5); increase of the interaction constant leads to increase in "viscosity" (or dissipation measure) of the contact environment, and this reduces the transfer probability for infectious carrier ("transition" from curve «1» to curve «3», Fig. 5). Also, in Fig. 5 one can observe the "blockage" effect. The "blockage" effect for the infectious diseases transfer has been demonstrated in the case when peak values (in the population density) are equal to each other (asymmetry parameter equals to 1); and under condition when "social temperature" is low. Existence of such effect essentially depends on the environment "activity" (social and prophylactic).

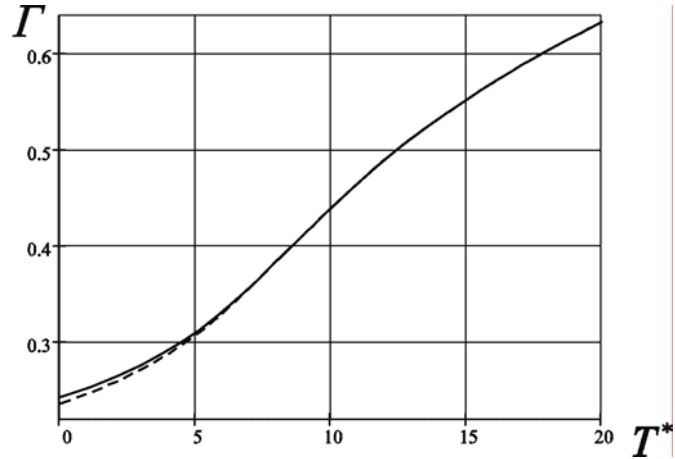

**Fig. 4.** Dependence of the transfer probability $\Gamma$ (for infectious carrier) on "social temperature" $T^*$.

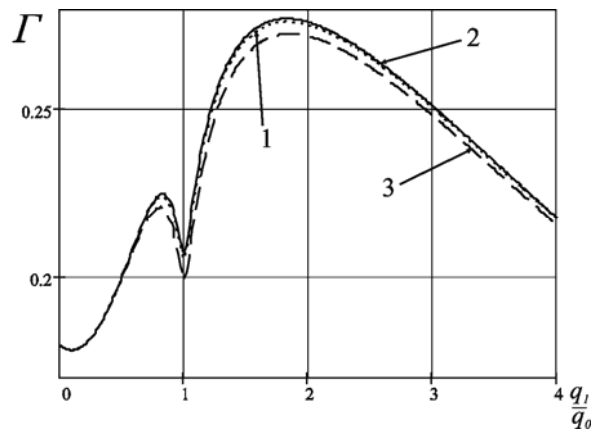

**Fig. 5.** Dependence of the transfer probability $\Gamma$ (for infectious carrier) on the asymmetry parameter $\dfrac{q_1}{q_0}$ (for population density profile).

In summary, in this paper applicability of the instanton approach to modeling of the infectious diseases transfer dynamics has been shown on an example of tuberculosis. Controllable features of influence on the infectious diseases spread rate have been revealed. In case of "blockage" effect, influence of the slow-changeable factors (such as the population density profile) has been accounted for; also we have shown the role of such dynamical factors as "social temperature" and other parameters of the infectious transfer environment in the infectious transfer control.

**Acknowledgements.** This work has been partly supported by grants: JSPS grant ID N S-05052; grant "Iryokikicenter", "Nano Medicine Supporting Program". The authors are grateful to Prof. K. Suzuki for the support of this work.